\documentclass[sigchi]{acmart}
\usepackage{subcaption}
\usepackage{float}

\AtBeginDocument{%
  \providecommand\BibTeX{{%
    \normalfont B\kern-0.5em{\scshape i\kern-0.25em b}\kern-0.8em\TeX}}}

\copyrightyear{2021}
\acmYear{2021}
\setcopyright{acmcopyright}\acmConference[preprint]{}
\acmDOI{0000/00000}
\begin{document}
\fancyhead{}
\title{Towards Automated Fatigue Assessment using Wearable \\Sensing and Mixed-Effects Models}

\author{Yang Bai$^{1}$, Yu Guan$^{1}$, Jian Qing Shi$^{2}$, Wan-Fai Ng$^{3}$} 
\affiliation{%
  \institution{$^{1}$Open Lab, Newcastle University, Newcastle upon Tyne, UK}
  \institution{$^{2}$Department of Statistics and Data Science, Southern University of Science and Technology}
  \institution{$^{3}$Translational and Clinical Research Institute, Newcastle University, UK}
  \institution{\{y.bai13, yu.guan, wan-fai.ng\}@newcastle.ac.uk, shijq@sustech.edu.cn}
  \country{}
}

\renewcommand{\shortauthors}{Bai, et al.}

\begin{abstract}
Fatigue is a broad, multifactorial concept that includes the subjective perception of reduced physical and mental energy levels. It is also one of the key factors that strongly affect patients' health-related quality of life. 
To date, most fatigue assessment methods were based on self-reporting, which may suffer from many factors such as recall bias. 
To address this issue, in this work, we recorded multi-modal physiological data (including ECG, accelerometer, skin temperature and respiratory rate, as well as demographic information such as age, BMI) 
in free-living environments, and developed automated fatigue assessment models. 
Specifically, we extracted features from each modality, and employed the random forest-based mixed-effects models,
which can take advantage of the demographic information 
for improved performance.  
We conducted experiments on our collected dataset, and very promising preliminary results were achieved. 
Our results suggested ECG played an important role in the fatigue assessment tasks.
\end{abstract}



\maketitle

{\fontsize{8pt}{8pt} \selectfont
\textbf{ACM Reference Format:}\\
Yang Bai, Yu Guan, Jian Qing Shi, Wan-Fai Ng. 2021. Towards Automated Fatigue Assessment using Wearable Sensing and Mixed-Effects Models. In \textit{Preprint.} ACM, New York, NY, USA, 3 pages. \url{https://doi.org/10.1145/nnnnnnn.nnnnnnn}}

\section{Introduction}

Fatigue is a broad, multifactorial concept that includes the subjective perception of reduced physical and mental energy levels. Most traditional fatigue assessments were based on self-reporting \citep{fatigue2003psychometric}.
Yet such subjective measurement has key limitations such as recall bias \cite{recall_bias}, and is challenging to quantify in a repeatable and reproducible way ~\cite{hardrepeat,aryal2017monitoring}. 
Although its reliability 
can be improved via tracking in long-terms or with high-frequency during short periods, the costs and patient burden can be increased significantly. 
Recently, wearable sensing and machine learning technologies have been used for automatic health assessment ~\cite{Shane_iswc, speech_health,autism, Tang_stroke, PD_1, PD_2, baby_stroke, chen2020automated, iranfar2021relearn, tag2019continuous}.
Through modelling the collected behaviour or physiological signals, health can be assessed in an objective and continuous manner, in contrast to the subjective, and non-continuous self-reporting methods. 
For automated fatigue assessment, potential sensing modalities include accelerometer\cite{Physical_fatigue_Jerk, physical_fatigue_HR_acc_Lasso}, HR\cite{fatigue_PPG, physical_fatigue_HR_acc_Lasso}, ECG\cite{mental_ECG}, EEG\cite{eeg2008}, EMG\cite{Physical_fatigue_muscle_exhaustion}, etc., 
based on which machine learning (ML) methods were used for modelling. 
Yet these approaches were tested on subjects in controlled environments.
Recently, deep learning (DL) methods have been introduced into fatigue assessment scenarios in ubiquitous and wearable computing.
In \cite{luo2020assessment}, self-supervised learning was proposed to extract informative features to explore the relationship between self-reported non-pathological fatigue and multimodal sensor data. 
In \cite{bai2020fatigue}, several sequential DL models including self-attention LSTM were developed for fatigue assessment.
Despite the effectiveness of both DL approaches\cite{luo2020assessment}\cite{bai2020fatigue}, it is hard to understand the decision-making process, and they also ignored the heterogeneity nature among different subjects, such as age, weight, height. 
In this work, we collected physiological data (as well as demographic information) from 21 participants over 7 days in the free-living environment and conducted preliminary analysis based on random-forest mixed-effects models.
These models not only can adapt to personal factors (such as BMI, age) for improved performance but also can output feature importance, which is a practical solution for interpretable fatigue assessment. 


\section{Methodology}

\subsection{Data Acquisition}
In this fatigue assessment study, we collected data from 21 patients with fatigue, with the corresponding scores of the physical and mental fatigue subscale 
ranging from 16 to 20
in the Multidimensional Fatigue Scale (MFI)\cite{smets1995multidimensional}
\footnote{In additional to MFI, there are other popular alternatives including Psychomotor Vigilance Task (PVT)\cite{tag2019continuous}, which has not been applied in this study.}. 
The participants were asked to wear two medical-grade devices (for 7 days in free-living environments), namely, Actigraph GT9X Link \cite{ActiGraph} and Vital Patch \cite{vitalpatch}, from which sensor data such as Actigraphy, ECG, skin temperature and respiratory rate data can be acquired.
The visual analog scale questionnaires were also sent out to participants asking the question "How severe has your fatigue been now?". 
The fatigue score ranging from 0 to 10 represents fatigue levels from "No fatigue" to "Maximal imaginable fatigue".
During the 7-day data collection period, the subjects were asked to record their fatigue levels 4 times a day, i.e., morning(6am-12pm), afternoon(12pm-6pm), evening(6pm-12am), and night(12am-6am).
Accordingly, we divided daily physiological signals into 4 segments, corresponding to the 4 daily recorded fatigue scores.

\subsection{Preprocessing and Feature Engineering}
During data collection, it is inevitable to face noisy or missing data issues  caused by device connection problems, inappropriate wearing, non-wearing, etc.
After removing the irrelevant and noisy information from raw signals, feature engineering can be performed for each specific modality, namely, Actigraphy, ECG, skin temperature and respiratory rate.

\noindent\textbf{Modality-specific feature extraction.}
Within each 6-hour segment, we first divided the raw physiological signals that consist of ECG, Actigraphy, Skin temperate and respiratory rate data into 5-min windows. 
For each window, we derive the corresponding features from the 4 modalities. As for ECG and Actgraphy feature extraction, following \cite{bai2020fatigue} we extracted 30 HRV features and 8 Actigraphy features. 
As for skin temperature and respiratory rate signals, we calculate 10 statistical features for each modality, i.e., mean value, standard deviation, minimum value, maximum value, skewness and kurtosis, 25th percentile, 50th percentile, 75th percentile and maximum drop. Now the data dimension is $D \times T$ for a segment, where $D=30+8+10+10=58$ indicates the feature number from the four modalities, while $T$ is the total number of (the 5-min) windows inside each 6-hour segment (e.g., $T=72$ if without missing data).
After feature extraction, we have 422 data points from 21 patients,
where each data point represents a feature sequence $\boldsymbol{X} = \{\boldsymbol{x}_t\in \mathbb{R}^D \}^T_{t=1}$, and the corresponding fatigue score $y \in \{0,1,...,10\}$. 

\noindent\textbf{High-level feature extraction.}
We further extracted high-level features from the 
sequence $\boldsymbol{X}$ over the time axis. 
To be concrete, for each dimension, we derived 13 descriptive statistics, namely 10th percentile, 25th percentile, 50th percentile, 75th percentile, 90th percentile, mean, minimum, maximum, standard deviation, skewness and kurtosis, interquartile range (IQR) and energy. 


\subsection{Random Forest Mixed Effects Models}
With the extracted high-level features, we can build interpretable models such as random forest (RF) for fatigue score prediction. 
Specifically, the Random Forest Mixed Effects Models (MERF)\cite{hajjem2014mixed} were applied, which can also model the personal factors. 
To be concrete, MERF consists of a fixed effect part and a random effect part, where the fixed effect part can be modelled by RF while the random effect part can be modelled using demographic information.
During modelling process, we grouped data into clusters using age, body mass index (BMI), or both. After modelling, MERF can then output demographic-specific fatigue assessment results.

\section{Experiments}
To evaluate the the effectiveness of the MERF models, we performed 5-fold cross-validation
(5-fold-CV). Root Mean Square Error (RMSE), Mean Absolute Error (MAE), Mean Absolute Percentage Error (MAPE) and Correlation Coefficient were used to measure the performance.


\noindent\textbf{Evaluation Results.}
Table 1 shows the results of several approaches, including linear model\cite{bai2020fatigue}, Random forest, and MERF with different grouping criteria (by Age, by BMI and by Age$\&$BMI).
We can see random forest approaches yielded better results than linear model\cite{bai2020fatigue}, and mixed effects models can further reduce the errors.
Specifically, MERF clustered by Age$\&$BMI achieved the best performance (i.e., lowest error and highest correlation), indicating demographic information may play important roles in automated fatigue assessment and they should be considered during the modelling process. 

\begin{table} [h]
\centering
\caption{Model Comparison} 
\begin{tabular}{lrrrr}
\toprule

\multicolumn{1}{c}{Method} 

&
\multicolumn{1}{c}{RMSE} &
\multicolumn{1}{c}{MAE}   &
\multicolumn{1}{c}{MAPE}   &
\multicolumn{1}{c}{Corr}   \\ 

\midrule

Linear Regression\cite{bai2020fatigue}   & 2.66$\pm$0.39 & 2.08$\pm$0.27 & 0.59$\pm$0.11 & 0.41      \\
Random Forest & 1.98$\pm$0.08 & 1.56$\pm$0.07 & 0.47$\pm$0.05 & 0.71      \\
MERF Age & 1.82$\pm$0.10 & 1.38$\pm$0.08 & 0.38$\pm$0.04 & 0.74      \\
MERF BMI & 1.88$\pm$0.11 & 1.47$\pm$0.07 & 0.42$\pm$0.06 & 0.73      \\
MERF Age\&BMI& 1.78$\pm$0.13 & 1.35$\pm$0.09 & 0.36$\pm$0.07 & 0.75      \\

\bottomrule
\end{tabular}

\label{tab:ablation study}
\end{table}

\begin{figure}[h]
\begin{center}
    \includegraphics[width=0.65\linewidth]{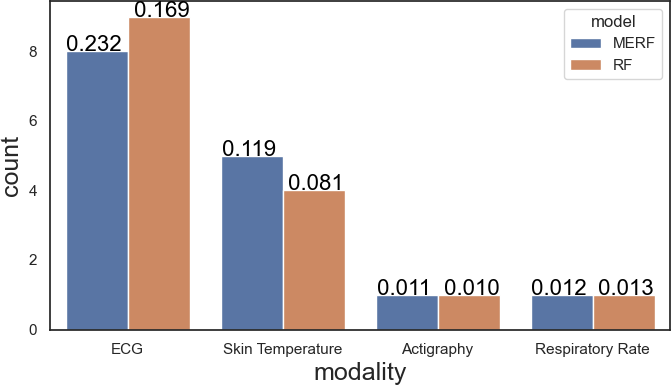}
\end{center}
\vspace{-0.4cm}
   \caption{Modality importance based on the top 15 features (for MERF and random forest); modality importance scores were showed on top of each bar}
\label{fig:f_imp}
\end{figure}

\noindent\textbf{Modality Importance.}
For random forest based approaches, it is straight-forward to output the feature importance (e.g., based on the aggregated information gains from the trees).  
Here for random forest and MERF, we also calculated the modality importance based on the top 15 features, and 
in Fig.\ref{fig:f_imp} we reported 
both the important feature counts (i.e., y-axis) and the corresponding aggregated importance scores (i.e.,scores on the bars) for each modality. 
We can see the most informative modality was ECG, which is inline with
existing medical studies ~\cite{fatigue_PPG, physical_fatigue_HR_acc_Lasso}. The least important modalities were accelerometer or respiratory rate, and they have the lowest feature counts or importance scores.

\section{Conclusion}
In this work, we developed an automated fatigue assessment system using wearable sensing and machine learning approaches. 
We collected multimodal data (ECG, accelerometer, skin temperature, respiratory rate) as well as demographic information(age, BMI) from 21 patients with fatigue symptoms. 
Specifically, we used MERF for the modelling, which can make the system adapt to the personal factors such as age and BMI. Promising results were achieved in this initial study, and 
we also found modality ECG contributed the most in the fatigue assessment tasks.


\begin{acks}

This project is supported by the Newcastle NIHR Biomedical Research Centre. 
The authors would also like to thank the support from SUSTech's student visiting scheme.


\end{acks}
\balance
\bibliographystyle{ACM-Reference-Format}
\bibliography{sample-base}

\end{document}